\documentclass[11pt,draft]{article}

\usepackage[latin1]{inputenc}

\usepackage{a4wide,amsfonts,dsfont,amssymb,rotating,wasysym,lscape,chngpage}

\usepackage{enumerate}

\usepackage{cmmib57}

\newtheorem{thm}{Theorem}[section]

\newtheorem{lem}{Lemma}[section]
\newtheorem{corollary}{Corollary}[section]
\newtheorem{observ}{Remark}[section]

\newtheorem{ejemrom}{Example}[section]
\newenvironment{proof}{\noindent {\it Proof:}\ }{$\square$}

\newenvironment{example}{\begin{ejemrom}\rm}{$\square$\end{ejemrom}}

\newcommand{\shapet}{{\mathbf{K}}}
\newcommand{\Lfour}{\mathbb{L}^4}
\newcommand{\R}{\mathbb{R}}
\newcommand{\G}{\mathbf{G}}
\newcommand{\ka}{\mathbf{k}}
\newcommand{\ele}{\mathbf{l}}
\newcommand{\arcsinh}{\mathrm{arcsinh}}

\def\dotsearrow{\hspace{-5mm}\dot{\hspace{5mm}\searrow}}
\def\dotswarrow{\hspace{5mm}\dot{\hspace{-5mm}\swarrow}}

\begin{document}

\title{Boost Invariant  Marginally Trapped Surfaces in Minkowski 4-Space}

\author{Stefan Haesen and Miguel Ortega\thanks{The
first author is partially supported by the Research Foundation -
Flanders project G.07.0432, and both authors are partially
supported by the MEC Grant MTM2007-60731 and the Junta de
Andaluc\'{\i}a Grant P06-FQM-01951.} }

\maketitle

\begin{abstract}
The extremal and partly marginally trapped surfaces in Minkowski
4-space, which are invariant under the group of boost isometries,
are classified. Moreover, it is shown that there do not exist
extremal surfaces of this kind with constant Gaussian curvature. A
procedure is given in order to construct a partly marginally
trapped surface by gluing two marginally trapped surfaces which
are invariant under the group of boost isometries. As an
application, a proper star-surface is constructed.
\end{abstract}


\section{Introduction}

In the study of the singularity theorems in general relativity,
trapped surfaces play a fundamental role. These surfaces,
introduced by Penrose \cite{penrose}, have the physical property
that the two null congruences normal to the surface are both
converging. From the mathematical point of view, the light
converging condition means that the mean curvature vector, which
measures the tension of the surface coming from the surrounding
space, is a timelike vector everywhere on the surface. If the mean
curvature vector is future- or past-pointing all over the surface,
the trapped surface is accordingly called future- or past-trapped.

The existence of closed trapped surfaces has been investigated in
several types of spacetimes. For example, in \cite{ellis,malec}
the formation of trapped surfaces in several cosmological is
studied, while in \cite{marssenovilla} the non-existence of these
closed surfaces is shown for strictly stationary spacetimes. In
\cite{dafermos} it is shown that the existence of one trapped
surface in a spherically symmetric spacetime is sufficient to
ensure the formation of a black hole and the completeness of null
infinity. In \cite{senovilla} a generalization of the concept of
trapped surface is given for codimension 2 submanifolds of a
Lorentzian space.

If the condition on the mean curvature vector is relaxed to be a
causal vector on the surface, the surface is called nearly trapped
and if the mean curvature vector is every lightlike or null, it is
called marginally trapped. In the mathematics literature, these
surfaces are called quasi-minimal and were studied in, e.g.,
\cite{rosca}. Although for their use in the singularity theorems
it is crucial that the marginally trapped surfaces are compact, in
recent work this condition is often not assumed. For example, in
\cite{chen,chen2} marginally trapped surfaces are studied in
Lorentzian space forms and Robertson-Walker spaces with an extra
assumption on the second fundamental form.

A complete classification of spacelike surfaces in a 4-dimensional
Lorentzian spacetime, containing the above cases, was recently
given in \cite{senovilla2}. In the following we make reference to
the notation introduced there.

In this text we consider spacelike surfaces in the Minkowski space
$\Lfour$ which are invariant under the following subgroup of
direct, linear isometries of $\Lfour$,
\[
\mathbf{G}   =  \left\{ B_{\theta} =\left( \begin{array}{cccc} \cosh(\theta) & \sinh(\theta) & 0 & 0 \\ \sinh(\theta) & \cosh(\theta) & 0 & 0 \\ 0 & 0 & 1 & 0 \\
0 & 0 & 0 & 1  \end{array} \right) \, : \, \theta\in\R\right\},
\]
well-known as \textit{boost} isometries. We also recall that they preserve the natural timelike orientation of $\Lfour$. We obtain the following results:
\begin{enumerate}

\item In Theorem \ref{marginales}, we classify all $\G$-invariant partly
marginally trapped surfaces in $\Lfour$, including a rigidity
result. The rest of this paper is a collection of corollaries of
this theorem and some examples. In particular, we  exhibit how to
construct examples of marginally trapped surfaces which were not
covered by the classification in \cite{chen}. In fact, the
condition of positive relative nullity assumed there can be
geometrically interpreted as saying that the intrinsic and
extrinsic curvature of the surface are the same. Hence, only the
surfaces of Corollary~\ref{flat} are covered by the classification
in \cite{chen}.

\item In Corollary \ref{extremal}, we classify all $\G$-invariant extremal surfaces in $\Lfour$, showing that they are contained in a totally geodesic  $\mathds{L}^3$.

\item In Section \ref{sec:glue} a general procedure is given to construct
$\G$-invariant partly marginally trapped surfaces by gluing two
$\G$-invariant marginally trapped surfaces. This method can be easily extended to a countable family of such surfaces. As an application, in
Example \ref{<punto>}, a null future-trapped surface is
constructed, while in Example \ref{ex:star} a method is given to
construct various $\varhexstar$-surfaces in Minkowski space. Recall \cite{senovilla2} for the definition of $\varhexstar$-surfaces.

\item Since the Gaussian curvature $K$ is invariant along the orbits
of the action of $\G$, the function $K$ only depends  on one
parameter. In this way, in Corollary \ref{GaussKR2}, we show that
it is possible to construct a $\G$-invariant spacelike surface
with prescribed Gaussian curvature. In addition, the surface might
be extremal or marginally trapped, according to a condition on the
profile curve. Also, it is possible that the time-orientation of
the mean vector field may vary from future to past (or viceversa),
on different regions of the surface where it is marginally
trapped. In such case, the  boundary of two such regions must be
of extremal points. In other words, the surface is an example of a
$\varhexstar$-surface. Moreover, we construct an example of a
$\G$-invariant marginally trapped surface with non-constant
bounded Gaussian curvature.

\item In Corollary \ref{no-max}, we show the non-existence of $\G$-invariant extremal surfaces in $\Lfour$ with constant Gaussian curvature.

\item In Corollary \ref{flat}, we describe the $\G$-invariant flat
marginally trapped surfaces in $\Lfour$. An open problem remains
to find  the $\G$-invariant marginally trapped surfaces with
constant Gaussian curvature. In Example \ref{K<0}, we give a
family of surfaces with negative constant Gaussian curvature. In
Example \ref{K>0}, an example of a surface with constant Gaussian
curvature $K=1$ is given, but by using transcendental functions,
since the desired primitives cannot be explicitly computed.
\end{enumerate}


\section{Set up}

Let $\Lfour$ be the 4-dimensional Lorentz-Minkowski space with the
flat metric given in local coordinates by
$$ g_o=-\mbox{d}x_1^2 + \mbox{d}x_2^2 +\mbox{d}x_3^2 +\mbox{d}x_4^2.$$
Given a connected immersed surface $S$ in $\Lfour$, we call $g$
the restriction of the metric $g_o$ to $S$. Also, we assume that
the metric $g$ is positive-definite, i.~e., the surface is
space-like. Let $\nabla$ be the Levi-Civita connection of
$(\Lfour,g_o)$. Given two smooth vector fields  ${X}$, ${Y}$
tangent to $S$, we denote the normal component of
$\nabla_{{X}}{Y}$ by $\shapet({X},{Y})$, where
$$ \shapet:\mathfrak{X}(S)\times \mathfrak{X}(S)\longrightarrow \mathfrak{X}^{\perp}(S),$$
which is usually called  the \textit{shape tensor} of $S$ in
$\Lfour$. Given a normal vector  ${\eta}$, the \textit{shape
operator} $A_{\eta}$ associated with $\eta$  is the endomorphism
of $TS$ given by $g_o(\shapet({X},{Y}),\eta)=g(A_{\eta}{X},{Y})$,
for any tangent vectors ${X}$, ${Y}$ to $S$. The \textit{mean
curvature vector} $H$ is the trace of the shape tensor,
$$ H=\frac{1}{2}\mathrm{tr}_{g} \shapet \in \mathfrak{X}^{\perp}(S).$$
The component of $H$ along a given normal direction $\eta$ is
called the \textit{expansion along} $\eta$,
$g(H,\eta)=\mathrm{tr}_g(A_\eta)$.

\smallskip

A surface $S$ in $\Lfour$ is \textit{boost invariant} if it is
invariant by the group $\G$, i. e., if $B_{\theta}(S)=S$ for any
$\theta\in\R$. Hence, the spacelike surface $S$ has to lie in one
of the two $\G$-invariant regions $\mathcal{R} =
\{(x_{1},x_{2},x_{3},x_{4})\in\Lfour:\, \mid\!x_{1}\!\mid >
\mid\!x_{2}\!\mid, x_{1}>0\}$, or the one with $x_{1}<0$,
respectively. In the following we will always assume
$S\subset\mathcal{R}$, since the other case is analogous. Since
the set of fixed points of $\G$ is $\{(x_1,x_2,x_3,x_4)\in\Lfour :
x_1=x_2=0\}$, we can introduce a parametrization $X(s,\theta)$ on
a dense open subset $\Sigma_{\alpha}$  of $S$ as follows,

\[ \Sigma_{\alpha} = \{ X(s,\theta)=\alpha(s)\cdot B_{\theta}\, : \, s\in I,
\theta\in\R\}, \]

{\noindent}where

\[ \alpha:I\subset\mathbb{R}\longrightarrow \mathcal{P} = \{(x_1,x_2,x_3,x_4)\in\Lfour : x_2=0, x_{1}>0\}, \
\alpha(s)=(\alpha_1(s),0,\alpha_3(s),\alpha_4(s)). \]

{\noindent}Also, since the surface $\Sigma_{\alpha}$ has to be space-like (also called
Riemannian), the curve $\alpha$ should be space-like, i.~e.,
\begin{equation}
-(\alpha_1')^2+(\alpha_3')^2 + (\alpha_{4}^{\prime})^{2}>0.
\end{equation}

We consider a local orthonormal basis $\{\eta_1,\eta_2\}$ of the
normal bundle of $\Sigma_{\alpha}$ in $\Lfour$, where $\eta_1$ is
future-pointing time-like and $\eta_2$ is space-like.  If we
denote by $A_i$ the shape operator associated with $\eta_i$,
$i=1,2$, the shape tensor can be written as
$$ \shapet( X, Y)=-g(A_1 X, Y)\eta_1+g(A_2 X, Y)\eta_2,$$
for any tangent vector fields $ X, Y$ to $\Sigma_{\alpha}.$  From
the classical theory of surfaces (see e.g. \cite{struik}), with
the notation
\begin{equation}
\begin{array}{c}
E=g(X_s,X_s),\ F=g(X_s,X_{\theta}), \ G=g(X_{\theta},X_{\theta}), \\ \vspace{0.5em}
e_i=g_o(X_{ss},\eta_i), \ f_i=g_o(X_{s\theta},\eta_i), g_i=g_o(X_{\theta\theta},\eta_i), \\ \vspace{0.5em}
h_i=\mathrm{tr}_g (A_i),
\end{array}
\end{equation}
\noindent we obtain
\begin{equation}
h_i=\frac{e_iG-2f_iF+g_iE}{EG-F^2}, \ i=1,2.
\end{equation}
Also, another useful local  basis $\{\ele,\ka\}$ of the normal bundle
of $\Sigma_{\alpha}$ in $\Lfour$ can be chosen such that they are
null, future-pointing and satisfying the normalization condition
$g_o( \ka, \ele)=-1$. The basis can be changed by choosing a positive
function defined on $S$, $\sigma$, in the following way,
$$  \ele \longrightarrow  \ele'=\sigma  \ele, \quad  \ka\longrightarrow  \ka'= \sigma^{-1}\ka.$$
In this note, we choose
$$  \ele =\frac{1}{\sqrt{2}}(\eta_1+\eta_2), \quad \ka =\frac{1}{\sqrt{2}}(\eta_1-\eta_2).$$
Now, the mean curvature vector field $H$ of $\Sigma_{\alpha}$
becomes
\begin{equation}H =\frac{1}{2}\Big( -h_1\eta_1+h_2\eta_2\Big)
= - \frac{\sqrt{2}}{4}(h_1-h_2) \ele -\frac{\sqrt{2}}{4}(h_1+h_2) \ka.
 \end{equation}
In particular, the \textit{expansions} along $\ele$ and $\ka$ are
given by
$$ \Theta_{\ele} = \frac{\sqrt{2}}{4}(h_1+h_2), \quad
\Theta_{\ka} = \frac{\sqrt{2}}{4}(h_1-h_2).
$$

We recall that a surface $S$  in $\Lfour$ is called {\em trapped}
if its mean curvature vector is everywhere time-like. If the mean
curvature vector is always light-like and non-zero in at least one
point, the surface is called {\em marginally trapped}. In case the
mean curvature vector is null on some region of the surface and
vanishes on another, the surface is called \emph{partly marginally
trapped}.

Besides the extrinsic mean curvature $H$, the intrinsic Gaussian
curvature $K$ of the surface can be expressed in terms of the
coefficients of the first and second fundamental forms as (see
e.g. \cite{struik}),
\begin{equation}\label{GaussianK}
K = \frac{-\det(A_{1}) + \det(A_{2})}{\det(g)} = \frac{ -e_{1}g_{1} +e_{2}g_{2} + f_{1}^{2}-f_{2}^{2}}{E G-F^{2}}\ .
\end{equation}


\section{Boost Invariant Surfaces}

We now consider a curve $\alpha:I\subset\R\rightarrow \mathcal{P}$
of the form

\[ \alpha(s) =
\Big(\alpha_{1}(s),\, 0,\, \alpha_{3}(s),\, \alpha_{4}(s)\Big). \]

{\noindent}The parametrization of $\Sigma_{\alpha}$, which is
invariant with respect to the rotation group $\G$, can be
written down as
\begin{equation} X(s,\theta) = \Big(\alpha_1(s)\cosh(\theta),\, \alpha_1(s)\sinh(\theta),\, \alpha_3(s),\, \alpha_4(s)\Big), \ s\in I, \theta\in\R.
\end{equation}
We assume that the derivative of the curve $\alpha$ has constant length, so
that $g_{o}(\alpha'(s),\alpha'(s))=c^2>0$ (for a suitable constant $c\in\R$). The derivatives of $X$ are
\begin{equation}
 X_s  =  (\alpha_1'\cosh(\theta),\alpha_1'\sinh(\theta),\alpha_3' , \alpha_4'), \quad
X_{\theta}  =  (\alpha_1\sinh(\theta),\alpha_1\cosh(\theta),0,0),
\end{equation}

{\noindent}and the Riemannian metric of the surface reads

\[ g = c^{2}\mbox{d}s^{2} + \alpha_{1}(s)^{2}\mbox{d}\theta^{2}\ . \]

{\noindent}A globally defined orthonormal tangent frame on
$\Sigma_{\alpha}$ is
\begin{equation}
u_1 = \frac{X_s}{c}, \quad u_2=\frac{X_{\theta}}{\alpha_1},
\end{equation}
and a globally defined  orthonormal basis of the normal bundle of
$\Sigma_{\alpha}$ is given by
\begin{eqnarray}
\eta_1&=& \frac{1}{c\sqrt{c^2+(\alpha_1')^2}} \Big(\cosh(\theta)
(c^2+(\alpha_1')^2),\, \sinh(\theta)
(c^2+(\alpha_1')^2),\,\alpha_1'\alpha_3',\,\alpha_1'\alpha_4'
\Big), \\
\eta_2&=&\frac{1}{\sqrt{c^2+(\alpha_1')^2}}
\Big(0,0,\,-\alpha_4',\,\alpha_3'\Big),
\end{eqnarray}
with $\eta_1$ future-pointing time-like and $\eta_2$ space-like. Moreover, the associated null basis is
\begin{eqnarray*}
\ele & = & \frac{1}{c \sqrt{2}\sqrt{c^2+(\alpha_1')^2}}
\Big(\cosh(\theta) (c^2+(\alpha_1')^2),\, \sinh(\theta)
(c^2+(\alpha_1')^2),\,\alpha_1'\alpha_3'-c\,
\alpha_4',\,\alpha_1'\alpha_4' +c\, \alpha_3' \Big) \\
 \ka & = & \frac{1}{c \sqrt{2}\sqrt{c^2+(\alpha_1')^2}}
\Big(\cosh(\theta) (c^2+(\alpha_1')^2),\, \sinh(\theta)
(c^2+(\alpha_1')^2),\,\alpha_1'\alpha_3'+c\,
\alpha_4',\,\alpha_1'\alpha_4'-c\, \alpha_3' \Big).
\end{eqnarray*}
A straightforward computation shows that
$$
e_1= \frac{-c \alpha_1''}{\sqrt{c^2+(\alpha_1')^2}}, \quad f_1=0, \quad
g_1=\frac{-\alpha_1}{c}\sqrt{c^2+(\alpha_1')^2},
$$
\begin{equation} \label{II-R2}
 e_2=  \frac{-\alpha_4'\alpha_3''+\alpha_4''\alpha_3'}{\sqrt{c^2+(\alpha_1')^2}}, \quad f_2=g_2=0,
\end{equation}
$$h_1=-\,\frac{c^2+(\alpha_1')^2+\alpha_1\alpha_1''}{c \alpha_1\sqrt{c^2+(\alpha_1')^2}}, \quad  h_2=\frac{-\alpha_4'\alpha_3''+\alpha_3'\alpha_4''}{c^2\sqrt{c^2+(\alpha_1')^2}}.
$$

{\noindent}Finally, the Gaussian curvature of a $\G$-invariant surface
$\Sigma_{\alpha}$ is

\begin{equation}\label{GaussK}
K = - \frac{\alpha_{1}^{\prime\prime}}{c^2\alpha_{1}}\ .
\label{gausscurvR2}
\end{equation}

{\noindent}Note that the surface $\Sigma_{\alpha}$ always lies on
the diagonal of Table 3 in \cite{senovilla2}. In the following we
are interested in the particular cases that the surface is
marginally trapped.

\section{Main Result}

We note that given a smooth function $\rho:I\longrightarrow\R$,
sometimes it is useful to modify it in such a way that it changes
its sign on one (or more) connected component of $\{s\in I:
\rho(s)\neq 0\}$, but it is still smooth after the modification.
We define the set of points $Z_{\rho}=\{s\in I: \rho(s)=0\}$.
Assume that there is a connected component $C$ of the interior
points of $Z_{\rho}$, and a boundary point $s_0\in\partial C$. By
the continuity of $\rho$, there is a real number $\delta>0$ such
that (i) for any other connected component $\tilde{C}$ of the
interior of $Z_{\rho}$, then $]s_0-\delta,s_0+\delta [ \cap
\tilde{C}=\emptyset$; and (ii) the set $]s_0-\delta,s_0+\delta [
\cap C$ is connected. Next, we can define a function
$\varepsilon:C\cup]s_0-\delta,s_0+\delta[\longrightarrow\{-1,0,1\}$
such that (i) if $s\in C$, then $\varepsilon(s)=0$; (ii)
$\varepsilon$ is constant $(\pm 1)$ on
$]s_0-\delta,s_0+\delta[\backslash C$; and (iii) $\varepsilon\rho$
is smooth on $C\cup]s_0-\delta,s_0+\delta[$. We should point out
that there might be more elements of $Z_{\rho}$ in
$]s_0-\delta,s_0+\delta[$, but there smoothness of $\rho$ yields
the choice of $\varepsilon$ as before. The choice of $\varepsilon$
can be extended to the whole domain of $\rho$ in such a way that
$\varepsilon\rho$ is smooth on $I$. This way, we have the
following lemma.

\begin{lem} \label{election} {\em Sign Choice}.
Given a smooth function $\rho:I\longrightarrow\R$, there exists
many (maybe only constant) functions
$\varepsilon:I\longrightarrow\{-1,0,1\}\subset\R$ such that
$\varepsilon\rho$ is smooth on $I$, and if $s\in I$ such that
$\rho(s)\neq 0$, then $\varepsilon(s)=\pm 1$.
\end{lem}
Usually, the simplest choice consists of setting $\varepsilon$ as
a constant on the whole domain of $\rho$. Different choices will
be useful in Section \ref{sec:glue}.

\begin{thm} \label{marginales} Given a constant $c>0$ and a smooth function $\alpha_1>0$, we define the function
$$ \rho =- \, \frac{c^2+(\alpha_1')^2+\alpha_1\alpha_1''}{\alpha_1}. $$
We choose a function $\varepsilon$ as in Lemma \ref{election}, such that $\varepsilon\rho$ is smooth. Next, we define the following functions
$$
\xi =\int \frac{ \varepsilon\, c\,
\rho}{c^2+(\alpha_1')^2}\mathrm{d}s,\quad  \alpha_3=\int \sqrt{c^2+(\alpha_1')^2}\,\cos(\xi)\mathrm{d}s, \ \mathrm{and} \
\alpha_4=\int \sqrt{c^2+(\alpha_1')^2}\,\sin(\xi)\mathrm{d}s.
$$
Then, the space-like curve
$\alpha(s)=(\alpha_1(s),0,\alpha_3(s),\alpha_4(s))$ is a profile
curve of a $\G$-invariant surface   $\Sigma_{\alpha}$ with
$$ H=\frac{\rho}{2c\sqrt{c^2+(\alpha_1')^2}} (-\eta_1+\varepsilon  \eta_2), \quad
g_{o}(\alpha',\alpha')=c^2.
$$
In addition,
\begin{enumerate}
 \item The surface $\Sigma_{\alpha}$ is marginally trapped on the points where $\rho\neq 0$. Moreover, $H$ is future-pointing if, and only if, $\rho<0$.
\item The surface $\Sigma_{\alpha}$ is extremal on the points where  $\rho =0$.
\item Given any other curve $\beta=(\beta_1,0,\beta_3,\beta_4)$ such that $g_{o}(\beta',\beta')=c^2$ and $\beta_1=\alpha_1$, for each connected component $C$ of $\{s\in I: \rho_{\alpha}(s)\neq 0\}$, there exists an affine isometry $F$ of $\Lfour$ such that $F(\Sigma_{\alpha\vert_{C}})=\Sigma_{\beta\vert_{C}}$.
\end{enumerate}
Conversely, any $\G$-invariant surface which is partly  marginally
trapped, admits a dense open subset of the form $\Sigma_{\alpha}$,
with $\alpha:I\longrightarrow \mathcal{P}$,
$g_{o}(\alpha',\alpha')=c^2$,
such that $\Sigma_{\alpha}$ can be constructed as above.
\end{thm}
\begin{proof} We first show the sufficient condition. Let $\alpha$ be a space-like
curve satisfying $g_{o}(\alpha',\alpha')=c^2>0$, with $c\in\R$ a
constant. We can rewrite this expression as
\begin{equation}\label{alfas-teta}
\alpha_3'=\sqrt{c^2+(\alpha_1')^2}\cos(\xi), \quad
\alpha_4'=\sqrt{c^2+(\alpha_1')^2}\sin(\xi),
\end{equation}
for a suitable smooth \textit{angle} function $\xi$. In this way,
up to integration constants, we can recover $\alpha_3$ and
$\alpha_4$ from the data $\xi$ and $\alpha_1$. Next, we assume
that $\alpha$ is the profile curve of a $\G$-invariant partly
marginally trapped surface. This means $\|H\|^2=0$. First, we
consider the case  $H\neq 0$. On a small enough open subset of the domain
of $\alpha$, there holds that $h_1=\varepsilon h_2$, with
$\varepsilon=\pm 1$. Thus, by Lemma \ref{election}, there exists a function $\varepsilon$ taking the values $\{-1,0,1\}$ such that $h_1=\varepsilon h_2$.  Bearing in mind (\ref{II-R2}), we see that this is equivalent to
\begin{equation}\label{h1=h2}
\rho  = -\, \frac{c^2+(\alpha_1')^2+\alpha_1\alpha_1''}{\alpha_1} =
\varepsilon \frac{-\alpha_4'\alpha_3''+\alpha_3'\alpha_4''}{c},
\end{equation}
where $\rho$ is a smooth function. We should point out that when $H=0$, then $h_1=h_2=0$, so that we can choose $\rho =0$ (and $\varepsilon=0$) in such points. Next, by inserting  (\ref{alfas-teta}) in (\ref{h1=h2}), we obtain
\begin{equation}\label{tetaprime}
 \xi' = \frac{\varepsilon\, c\, \rho}{c^2+(\alpha_1')^2}.
\end{equation}
This expression allows one to recover $\xi$ from $\rho$ and
$\alpha_1$, up to an integration constant. Now, by (\ref{II-R2})
and (\ref{h1=h2}), and recalling $\rho=h_1 c
\sqrt{c^2+(\alpha_1')^2}=\varepsilon h_2c
\sqrt{c^2+(\alpha_1')^2}$, it readily follows that
$$H= \frac{\rho}{2c\sqrt{c^2+(\alpha_1')^2}} (-\eta_1+\varepsilon
\eta_2).$$
From this expression, since $\eta_1$ is
future-pointing, $H$ is future-pointing if, and only if, $\rho<0$.
It only remains to check the rigidity condition of item 3. Indeed,
if $\alpha_1=\beta_1$ and $g_{o}(\beta',\beta')=c^2$, both
functions $\alpha_1$ and $\beta_1$ determine the very same
function $\rho$. Let $C$ be a connected component of $\{s\in I: \rho_{\alpha}(s)\neq 0\}$, so we work on $C$. Next, if $\xi_{\alpha}$ and $\xi_{\beta}$ are
the corresponding \textit{angle} functions, then
$\xi_{\alpha}'=\delta \xi_{\beta}'$, for a suitable constant
$\delta=\pm 1$, which depends on the functions $\varepsilon$
associated with $\alpha$ and $\beta$. This way,
$\xi_{\alpha}=\delta\xi_{\beta}+\xi_0$, where $\xi_0$ is just an
integration constant. An elementary computation shows
$$ (\beta_3',\beta_4') = (\alpha_3',\alpha_4')\left( \begin{array}{cc} 1 & 0 \\ 0 & \delta \end{array} \right)
\left( \begin{array}{cc} \cos(\xi_0) & \sin(\xi_0) \\ -\sin(\xi_0) & \cos(\xi_0) \end{array} \right).
$$
Next, if the integration constants are $\alpha_{3}^0$ and $\alpha_{4}^0$, we call $v= (0,0,\alpha_{3}^0,\alpha_{4}^0)$. It is clear that the affine isometry $F:\Lfour\longrightarrow\Lfour$,
$$
F(x_1,x_2,x_3,x_4)=(x_1,x_2,x_3,x_4)
\left( \begin{array}{cccc} 1 & 0 & 0 & 0 \\ 0 & 1 & 0 & 0 \\ 0 & 0 & 1 & 0 \\ 0 & 0 & 0 & \delta \end{array} \right)
\left( \begin{array}{cccc} 1 & 0 & 0 & 0 \\ 0 & 1 & 0 & 0 \\ 0 & 0 & \cos(\xi_0) & \sin(\xi_0) \\
0 & 0 & -\sin(\xi_0) & \cos(\xi_0) \end{array} \right)+v,
$$
satisfies $F\circ \alpha\vert_{C}=\beta\vert_{C}$ and thus
$F(\Sigma_{\alpha\vert_{C}})=\Sigma_{\beta\vert_{C}}$.

Finally, the converse is a straightforward computation from equations  (\ref{alfas-teta}), (\ref{h1=h2}) and (\ref{tetaprime}).
\end{proof}

\begin{corollary}\label{extremal} Let $\Sigma_{\alpha}$ be a $\G$-invariant extremal surface. Then, a unit profile curve is given by
$$ \alpha(s)=\left(f(s), 0,
\cos(\xi_0) \sqrt{a_1} \arctan\Big(\frac{s+a_2}{f(s)}\Big),
\sin(\xi_0) \sqrt{a_1} \arctan\Big(\frac{s+a_2}{f(s)}\Big)
 \right),
$$
where $f(s)=\sqrt{a_1-(s+a_2)^2}$, and $a_1$, $a_2$, $\xi_0\in\R$, $a_1>0$, being integration constants. In particular, $\Sigma_{\alpha}$ is immersed in a totally geodesic Lorentzian 3-plane.
\end{corollary}

\begin{proof} If we assume  $H= 0$, we have $h_1=h_2=0$. By (\ref{II-R2}), with $c=1$, we obtain
\begin{equation}\label{las3}
 1+(\alpha_1')^2+\alpha_1\alpha_1''=0 \quad \mbox{and} \quad -\alpha_4'\alpha_3''+\alpha_3'\alpha_4''=0.
\end{equation}
Now, the general solution of (\ref{las3}) is
$\alpha_1(s)=\varepsilon\sqrt{a_1-(s+a_2)^2}$, where $a_1$,
$a_2\in\R$ are integration constants with $a_1>0$ and
$\varepsilon=\pm 1$. However, since we are always assuming
$\alpha_1$ to be positive, the case $\varepsilon=-1$ cannot occur.
Next, we make use of Theorem \ref{marginales}. This way, $\rho=0$,
which readily implies $\xi(s)=\xi_0$ for a certain constant
$\xi_0\in\R$. A straightforward computation gives the expressions
of $\alpha_3$ and $\alpha_4$.  In addition, the second equation of
(\ref{las3}) implies that the shape operator $A_2=0$.
Finally, we show that the surface is contained in a totally
geodesic  $\mathbb{L}^3$. To do so, we observe that the curve
$\alpha$ is contained in the Lorentzian 2-plane
$\{(x_1,x_2,x_3,x_4)\in\Lfour : x_2=0,
\sin(\xi_0)x_3=\cos(\xi_0)x_4\}$ and that the $\G$-orbits are
orthogonal to this plane. \end{proof}

\section{Gluing $\G$-Invariant Marginally Trapped Surfaces}
\label{sec:glue}

The aim of this section is to give a general method to construct a
$\G$-invariant partly marginally trapped surface from two
$\G$-invariant marginally trapped surfaces.

We recall that the following function $f:\R\longrightarrow\R$ is
smooth,
$$  f(s)= \left\{
\begin{array}{cc} \mathrm{e}^{-1/s}, & \mbox{ if } s>0 \\
0, & \mbox{ if } s\leq 0.\end{array} \right.$$ From it, it is
possible to construct $\tilde{f}(s)=\frac{f(s)}{f(s)+f(1-s)}$ and
$\bar{f}(s)=\tilde{f}(s+2)\tilde{f}(2-s)$, defined on $\R$, both
non-negative and smooth. But $\bar{f}$ satisfies that if $s\in
[-1,1]$, then $\bar{f}(s)=1$, and if $s\leq -2$ or $s\geq 2$, then
$\bar{f}(s)=0$. By simple changes of parameters, we obtain the
following lemma.

\begin{lem} \label{lem:lem1}
Let $p_1$, $p_2$, $q_1$ and $q_2 \in \R\cup\{-\infty,+\infty\}$ such that
$-\infty\leq p_1<q_1<p_2<q_2\leq +\infty$. Let  $d=\min\{q_1-p_1,p_2-q_1,q_2-p_2\}/4$.
Then, there exist three smooth functions
$f_i:]p_1,q_2[\longrightarrow \R$, $i=0,1,2$, satisfying the
following properties.
\begin{enumerate}
\item $0\leq f_i\leq 1$, for $i=0,1,2$.
 \item If $s\in]p_1,q_1-d]$, $f_1(s)=1$, and $f_1(s)=0$ if $s\geq q_1$.
\item If $s\in[q_1,p_2]$, $f_0(s)=1$, and $f_0(s)=0$ if $s\geq p_2+d$ or $s\leq q_1-d$.
\item If $s\in[p_2+d,q_2[$, $f_2(s)=1$ and $f_2(s)=0$ if $s\leq p_2$.
\end{enumerate}
\end{lem}
Next, we would like to construct a new partly marginally trapped surface from two marginally trapped surfaces. We describe the method in several steps.
\begin{description}
 \item[Preliminaries.] Given two unit curves,
 $\alpha^i:]p_i,q_i[\longrightarrow \mathcal{P}$, $i=1,2$, such that $\Sigma_{\alpha^i}$
is a marginal\-ly trapped surface, with $-\infty\leq p_{1}<q_{1}<p_{2}<q_{2}\leq +\infty$.
Denote the associated functions by $\rho_i$, $\xi_i$ and
$\varepsilon_i$, $i=1,2$. We write
$\alpha^i=(\alpha^i_1,0,\alpha^i_3,\alpha^i_4)$, $i=1,2$ for the
profile curves and define the numbers $m_1=(p_2+q_1)/2$ and
$m_2=(p_2+2d-q_1)/2$. Using Corollary \ref{extremal}, we consider
the curve $\alpha^0:]q_1-d,p_2+d[ \longrightarrow\mathcal{P}$
which is the profile curve of an extremal surface
$\Sigma_{\alpha^0}$, with $f(s)=\sqrt{m_2^2-(s+m_1)^2}$ and
$\xi_0=0$.

\item[Step 1.] Using Lemma \ref{lem:lem1}, we define the function
$\beta_1:]p_1,q_2[\longrightarrow\R$ as
$\beta_1=\sum_{i=0}^2f_i\alpha^i_{1}$, with the functions $f_{i}$
from Lemma \ref{lem:lem1}. Clearly, this function is smooth and
positive. Moreover, $\beta_1$ restricted to $]p_1,q_1-d[$,
$]q_1,p_2[$ and $]p_2+d,q_2[$, respectively, is equal to
$\alpha^1_1$, $\alpha^0_1$ and $\alpha_1^2$, restricted to the
respective intervals.

\item[Step 2.] Define the smooth function $\rho$ as in Theorem \ref{marginales} for $\beta_1$.
Also, if $s\in ]p_1,q_1-d[$, then $\rho(s)=\rho_1(s)$; if
$s\in]q_1,p_2[$, then $\rho(s)=0$; and if $s\in ]p_2+d,q_2[$, then
$\rho(s)=\rho_2(s)$.

\item[Step 3.] Define the function $\varepsilon:]p_1,q_2[\longrightarrow\R$, given by
$$ \varepsilon(s)=\left\{ \begin{array}{ccc} \varepsilon_1 & \mathrm{ if } & s\in ]p_1,q_1], \\
0 & \mathrm{ if } & s\in ]q_1,p_2], \\  \varepsilon_2 & \mathrm{ if } & s\in ]p_2,q_2[.
\end{array}\right.$$
It is a simple matter to check that the function $\varepsilon\rho$ is smooth.

\item[Step 4.] Define $\xi = \int \frac{\varepsilon\rho}{1+(\beta_1')^2}\mathrm{d}s$,
and $\beta_3$, $\beta_4$ as in Theorem \ref{marginales}. Thus,  the unit curve
$\beta=(\beta_1,0,\beta_3,\beta_4)$ is the profile curve of a
partly marginally trapped surface.

\item[Step 5.] It only remains to show how the original surfaces are related with the new one. For this we resort to item 3 of Theorem \ref{marginales}. Thus, if $s\in ]p_1,q_1-d[$, then $\beta_1(s)=\alpha_1^1(s)$ and $\varepsilon(s)=\varepsilon_1$, and therefore, there exists a direct affine isometry $F_1:\Lfour\longrightarrow\Lfour$ such that $F_1\circ {\alpha^1}_{\vert  ]p_1,q_1-d[} = \beta_{\vert  ]p_1,q_1-d[}$. Similarly, there exists a direct affine isometry $F_2:\Lfour\longrightarrow\Lfour$ such that $F_2\circ {\alpha^2}_{\vert  ]p_2+d,q_2[} = \beta_{\vert  ]p_2+d,q_2[}$.
\end{description}
This method can be easily extended to a countable family of
$\G$-invariant partly marginally trapped surfaces, as it is shown
in the following corollary.

\begin{corollary} Let $\{\alpha^k:]a_k,b_k[\longrightarrow\mathcal{P} / k\in\mathcal{K}  \subset\mathbb{N}\}$ be a countable family of unit curves with $a_k$, $b_k\in\R$ for any $k\in\mathcal{K}$, such that each $\Sigma_{\alpha^k}$, $k\in\mathcal{K}$, is a $\G$-invariant  partly marginally trapped surface. Then, there exists a $\G$-invariant partly marginally trapped surface $\Sigma_{\alpha}$ such that for each $k\in\mathcal{K}$, there exists an open subset $C_k$ of the domain of $\alpha$, and an affine isometry $F_k$ of $\Lfour$ satisfying  $F_k(\Sigma_{\alpha^k}) = \Sigma_{\alpha\vert_{C_k}}$.
\end{corollary}

\section{The Gaussian Curvature of $\G$-Invariant Partly Marginally Trapped Surfaces}

\begin{corollary} \label{GaussKR2}
Let $\kappa:I\subset\mathbb{R}\longrightarrow\R$ be a smooth
function. Given $s_0\in I$, there exist $\delta>0$ and a unit
space-like curve
$\alpha:(s_0-\delta,s_0+\delta)\longrightarrow\mathcal{P}$ such
that $\alpha$ is a unit profile curve of the  $\G$-invariant
surface $\Sigma_{\alpha}$ whose Gaussian curvature satisfies
$K(s,\theta)=\kappa(s)$ for any $(s,\theta)\in
(s_0-\delta,s_0+\delta)\times \R$. In addition,
\begin{enumerate}
 \item The surface $\Sigma_{\alpha}$ is marginally trapped on
the points of $(s_0-\delta,s_0+\delta)$ where
$1+(\alpha_1')^2-\kappa \alpha_1^2\neq 0$. Moreover, the mean
curvature vector $H$ of the surface $\Sigma_{\alpha}$ is
future-pointing if, and only if,
$\kappa<\frac{1+(\alpha_1')^2}{\alpha_1^2}$.

\item The surface $\Sigma_{\alpha}$ is extremal on the points of $(s_0-\delta,s_0+\delta)$ where  $1+(\alpha_1')^2-\kappa \alpha_1^2= 0$.
\end{enumerate}
\end{corollary}
\begin{proof} Given a smooth function $\kappa:I\subset\mathbb{R}\longrightarrow\R$,
let $\alpha_1:(s_0-\delta,s_0+\delta)\longrightarrow\R$ be a
positive solution of the differential equation $\alpha_1''=-\kappa
\alpha_1$ (recall expression (\ref{gausscurvR2})). Now, we only
need to resort to Theorem \ref{marginales}, taking the constant
$\varepsilon=1$. The condition $\rho <0$ becomes now $0>\rho= -
(1+(\alpha_1')^2+\alpha_1\alpha_1'')/\alpha_1$. From here, it
immediately follows that
$\kappa<\frac{1+(\alpha_1')^2}{\alpha_1^2}$.
\end{proof}

\begin{corollary} \label{no-max} There are no $\G$-invariant extremal surfaces with constant Gaussian curvature in $\Lfour$.
\end{corollary}
\begin{proof} Let $M$ be an extremal $\G$-invariant surface
with constant Gaussian curvature in $\Lfour$. Then, there exists a
unit curve $\alpha$ such that an open subset of $M$ is
$\Sigma_{\alpha}$. According to Corollary \ref{GaussKR2}, equation
$(\alpha_1')^2=K \alpha_1^2-1$ holds on an open interval. This
implies $K>0$. Next, by (\ref{gausscurvR2}), there holds that
$\alpha_1(s)=a_1\cos(\sqrt{K} \, s+a_2)$, for suitable integration
constants $a_1,a_2\in\R$. However, for this $\alpha_1$, equation
$(\alpha_1')^2=K \alpha_1^2-1$ holds only for some isolated points,
which is a contradiction.
\end{proof}

\begin{corollary}\label{flat} A $\G$-invariant marginally trapped
surface $\Sigma_{\alpha}$ is flat if, and only if, its unit
profile curve is, up to translations and orientation,
$\alpha(s)=(\alpha_1(s),0,\alpha_3(s),\alpha_4(s)$, where
\begin{eqnarray*}
\alpha_1(s)&=&a_1s+a_2>0, \\
\alpha_3(s)&=&\frac{a_1s+a_2}{\sqrt{a_1^2+1}}
\left(
\varepsilon\sin\Big(\varepsilon\frac{\log(a_1 s+a_2)}{b_1}+\xi_0\Big) +
a_1\cos\Big(\varepsilon\frac{\log(a_1 s+a_2)}{b_1}+\xi_0\Big)  \right)\\
\alpha_4(s)&=&\frac{a_1s+a_2}{\sqrt{a_1^2+1}}
\left(
\varepsilon\cos\Big(\varepsilon\frac{\log(a_1 s+a_2)}{b_1}+\xi_0\Big)
-a_1\sin\Big(\varepsilon\frac{\log(a_1 s+a_2)}{b_1}+\xi_0\Big)
\right),
\end{eqnarray*}
$a_1,a_2,\xi_0\in\R$ being integration constants, $\varepsilon=\pm 1$ and
$$ H = - \frac{\sqrt{1+a_1^2}}{a_1s+a_2}(-\eta_1+\varepsilon \eta_2).$$
\end{corollary}
\begin{proof} It is sufficient to recall that the surface is flat if, and
only if, $K=0$, and by (\ref{GaussK}), it follows that the
function $\alpha_1$ can be expressed as $\alpha_1(s)=a_1 s+a_2$
for some integration constants $a_1,a_2\in\R$. Now, we only have
to resort to Theorem \ref{marginales}.
\end{proof}

\section{Examples}

\begin{example} We consider the unit space-like curve
$$ \alpha:\R\longrightarrow\Lfour, \
\alpha(s)=\left( 1+\frac{s^2}{4}, 0, -3s+8\arctan(s/2), 4\log\Big(1+\frac{s^2}{4}\Big)-\frac{s^2}{4}\right).
$$
Let $\Sigma_{\alpha}$ be the $\G$-invariant surface whose profile
curve is $\alpha$. Then, this surface is pure, future-pointing,
marginally trapped, with
$$ H=\frac{3}{\sqrt{2}\sqrt{4+s^2}}\ele, \quad K=\frac{-2}{4+s^2}.$$
This is done by putting $\rho=-3/2$, $c=1$ and $\varepsilon=-1$ in
Theorem \ref{marginales}, and letting all integration constants be
zero.
\end{example}

\begin{example}\label{K<0} Given $c\in\R$, $c>0$, the spacelike curve
$\alpha:\R\longrightarrow\Lfour$, given by
\[
\alpha(s)=\left(  c \cosh(s), 0,
4 c \arctan(\tanh(s/2))-c \sinh(s), 2c \log(\cosh(s)) \right), \quad
g_{o}(\alpha',\alpha')=c^2,
\]
is a profile curve of a $\G$-invariant, future-pointing,
marginally trapped surface $\Sigma_{\alpha}$ with constant
Gaussian curvature $K=-1/c^2$ and $H=\frac{-2\sqrt{2}}{c}\ele$. In
fact, this example has been computed by choosing
$\alpha_1(s)=c\cosh(s)$, $c>0$, $\varepsilon=-1$ and making use of
Theorem \ref{marginales}.
\end{example}

\begin{example} \label{K>0} We consider the  unit curve $\alpha:]0,\pi[\longrightarrow\mathcal{P}$ given by
$$\alpha_1(s)=\sin(s), \ \alpha_3(s)=\int_s \sqrt{1+\cos^2(u)} \cos (\xi(u))\mathrm{d}u, \
\alpha_3(s)=\int_s \sqrt{1+\cos^2(u)} \sin (\xi(u))\mathrm{d}u,$$
where $\xi(u)=\arctan(\cos(u))+\log(\tan(u/2))$. According to
Theorem \ref{marginales}, the $\G$-invariant surface generated by
$\alpha$ is partly marginally trapped. Indeed, $\rho(s)=-
\frac{2\cos^2(s)}{\sin(s)}\leq 0$ and the constant $\varepsilon$
takes the value $\varepsilon=-1$. This surface is not (pure)
marginally trapped because $\rho(\pi/2)=0$. Moreover, by
(\ref{GaussK}), we see
$$ K=1, \quad H=\frac{\sqrt{2}\cos^2(s)}{\sin(s)\sqrt{1+\cos^{2}(s)}} \ele.$$
\end{example}

\begin{example} \label{<punto>} In this example, we use the procedure outlined
in Section \ref{sec:glue} to construct a unit profile curve of a
partly marginally trapped surface such that the mean curvature
vector is null, future-pointing on one region, zero on a second
region with non-empty interior and null, future-pointing on a
third region. In \cite{senovilla2} these surfaces are called null
future-trapped and denoted by
\begin{sideways}\begin{sideways}$\swarrow\dotsearrow$\end{sideways}\end{sideways}. For
the sake of simplicity, we let all the integration constants be
zero.

First, we consider two unit curves $\alpha,\beta:\R\longrightarrow
\mathcal{P}$,
\begin{eqnarray*}
\alpha(s) & = & (s^2+1, 0, -s\sqrt{s^2+1}+2\arcsinh(s), -3 \sqrt{s^2+1}), \\
\beta(s) & = & (s^2+1, 0, -s\sqrt{s^2+1}+2\arcsinh(s), 3 \sqrt{s^2+1}).
\end{eqnarray*}
Next, we choose a unit curve that generates an extremal surface,
$\varphi:]-2,2[\longrightarrow \mathcal{P}$,
$$
\varphi(s)=\left(\sqrt{4-s^2}, 0,
2\arcsin\left(\frac{s}{2}\right),0\right).
$$
A straightforward computation shows that, according to Theorem
\ref{marginales}, the associated $\rho$ functions are
$$ \rho_{\alpha}(s)= - \frac{6s^2+3}{s^2+1}= \rho_{\beta}(s), \quad \rho_{\varphi}(s)=0.
$$
Since $\rho_{\alpha}$ and $\rho_{\beta}$ are negative, both surfaces $\Sigma_{\alpha}$ and $\Sigma_{\beta}$ are future-pointing. Further, the associated $\varepsilon$ functions are taken to be the constants $\varepsilon_{\alpha}=1$ and $\varepsilon_{\beta}=-1.$ In order to recall the procedure of Section \ref{sec:glue}, we restric the curves to
$$\alpha:]-\infty,-1[\longrightarrow\mathcal{P}, \quad \beta:]1,+\infty[\longrightarrow\mathcal{P}.$$
\end{example}

\begin{example} \label{ex:star}
A $\varhexstar$-surface can be constructed by choosing the unit
curves $\alpha:]3,+\infty[\rightarrow\mathcal{P}$,

$$ \alpha(s) = (s^2+1, 0, -s\sqrt{s^2+1}+2\arcsinh(s), -3
\sqrt{s^2+1}), $$

{\noindent}and $\beta:]-\frac{3}{2},2[\rightarrow\mathcal{P}$,

$$ \beta(s) =
\left(\sqrt{-\frac{3}{2}s^{2}+2s+2},0,\beta_{3}(s),\beta_{4}(s)\right),
$$

{\noindent}with

$$ \beta_{3}(s) = \int
\sqrt{\frac{3s^{2}-4s+12}{-6s^{2}+8s+8}}\cos(\xi(s))\mbox{d}s,
\quad \beta_{4}(s) = \int
\sqrt{\frac{3s^{2}-4s+12}{-6s^{2}+8s+8}}\sin(\xi(s))\mbox{d}s, $$

{\noindent}and

$$ \xi(s) =
-\frac{\sqrt{6}}{3}\arcsin\left(\frac{3s-2}{4}\right)
+\arctan\left(\frac{-3s+2}{\sqrt{-6s^{2}+8s+8}}\right).
$$

{\noindent}Both curves are the profile curves of a future and past
marginally trapped surface, respectively.
Then, the procedure from Section \ref{sec:glue} gives different surfaces according to the chosen values of $\varepsilon_{\alpha}$ and $\varepsilon_{\beta}$, i.~e.,
\begin{center}
\begin{tabular}[t]{|c|c|c|c|} \hline
\multicolumn{2}{|c|}{null untrapped} & \multicolumn{2}{c|}{null dual} \\
\hline
\begin{sideways}$\swarrow\dotsearrow$\end{sideways} & \begin{sideways}\begin{sideways}\begin{sideways}$\swarrow\dotsearrow$
\end{sideways}\end{sideways}\end{sideways}
&
\raisebox{1.25em}{$\nwarrow$} $\dotsearrow$ &
$\dotswarrow$ \raisebox{1.25em}{$\nearrow$}
 \\
$\varepsilon_{\alpha}=-1$ & $\varepsilon_{\alpha}=1$ & $\varepsilon_{\alpha}=1$ & $\varepsilon_{\alpha}=-1$  \\
$\varepsilon_{\beta}=-1$ & $\varepsilon_{\beta}=1$ & $\varepsilon_{\beta}=-1$ & $\varepsilon_{\beta}=1$  \\
\hline
\end{tabular}
\end{center}
Moreover, if we also consider
$\gamma:]-\infty,-3[\longrightarrow\mathcal{P}$, $\gamma(s)$ with
the same expression as $\alpha(s)$, by iterating the procedure, it
is possible to obtain a surface of type \raisebox{-0.6em}{
\setlength{\tabcolsep}{0.2em}
\begin{tabular}[b]{rl}
$\nwarrow$ &  $\nearrow $ \\ &
$\dotsearrow$ \\
\end{tabular}}
\end{example}

\section{Conclusions}

Among the partly marginally trapped surfaces in Minkowski 4-space,
we are interested in the study of those which are invariant by
\textit{boost} isometries. As a main result, we are able to
classify them in Theorem \ref{marginales}. From this, we obtain a
fairly long list of corollaries. First, we obtain a classification
of boost invariant  extremal surfaces. Next, a careful reading of
Theorem \ref{marginales} gives rise to a method to construct
partly marginally trapped surfaces in Minkowski 4-space which may
include regions which are future or past-pointing, \textit{as
desired}, and extremal regions in between.  As an application, we
obtain  examples of proper $\varhexstar$-surfaces. Also, we show
that it is possible to construct a boost invariant surface with
prescribed Gaussian curvature. In particular, we show the
non-existence of boost invariant, extremal surfaces with constant
Gaussian curvature. In the text we make use of the notation
introduced in \cite{senovilla2}.

These methods may lead to the study of boost invariant generalized
horizons in Minkowski 4-space, since they are hypersurfaces
foliated by marginally trapped surfaces. These techniques may also
be applied to the study of marginally trapped surfaces which are
invariant by other subgroups of isometries in the Minkowski
4-space, as well as in other space-times.


\bibliographystyle{plain}

\vspace{10mm}

Stefan Haesen

Department of Mathematics

Katholieke Universiteit Leuven

Celestijnenlaan 200B

3001 Leuven, Belgium

E-mail: \texttt{Stefan.Haesen@wis.kuleuven.be}

\vspace{5mm}

Miguel Ortega

Departamento de Geometr\'{\i}a y Topolog\'{\i}a

Universidad de Granada

18071 Granada, Spain

E-mail: \texttt{miortega@ugr.es}

\end{document}